\documentstyle[11pt,twocolumn,epsf]{article}

\begin{document}
\title{\bf{Inverse versus Normal NiAs Structures as High Pressure Phases of FeO and MnO \\}}
\author{Z. Fang$^{*}$, K. Terakura$^{\dag}$, H. Sawada$^{*}$, T. Miyazaki$^{\ddag}$ \& I. Solovyev$^{*}$\\ \\
\small\it $^*$JRCAT, Angstrom Technology Partnership, 1-1-4 Higashi, Tsukuba, Ibaraki 305, Japan; \\
\small\it $^{\dag}$JRCAT, National Institute for Advanced Interdisciplinary Research, 1-1-4 Higashi, Tsukuba, Ibaraki 305, Japan; \\
\small\it $^{\ddag}$National Research Institute for Metals, 1-2-1 Sengen, Tsukuba, Ibaraki 305, Japan;}
\date{}
\maketitle
\thispagestyle{empty}

\bf Intensive high-pressure experiments on the transition-metal monoxides
have revealed that
FeO~\cite{FeO1,FeO2,FeO3,Mao,Fei} and MnO~\cite{MnO1,Noguchi,Kondo} undergo
pressure-induced phase transition. The high-pressure phase
of FeO was identified as the NiAs (B8) type~\cite{Mao,Fei},
while that of MnO is yet unclear~\cite{Noguchi,Kondo}.
The present theoretical study
predicts that the high-pressure phase of MnO is a metallic normal B8
structure (nB8), while that of FeO should take the inverse B8 structure (iB8).
The novel feature of the unique high-pressure phase of stoichiometric
FeO is that the system should be a band
insulator in the ordered antiferromagnetic (AF) state and that
the existence of a band gap leads to special stability of the phase.
The observed metallicity~\cite{metal1,metal2} of the high-pressure and high-temperature
phase of FeO may be caused by the loss of AF order and also by the
itinerant carriers created by non-stoichiometry.  
Analysis of x-ray diffraction experiments provides a further support to the
present theoretical prediction for both FeO and MnO. Strong stability
of the high-pressure phase of FeO will imply possible important roles in Earth's core.

\rm FeO and MnO at normal pressure are
typical examples of the Mott insulator~\cite{Mott} and their basic properties
are governed by the electron correlation whose strength is measured
by $U/W$ with $U$ the effective Coulomb interaction
integral between $d$ electrons and $W$ the $d$ band width.
As the $d$ band width is most directly controlled by pressure, the high-pressure studies 
have been regarded as a useful way to understand the
basic properties of Mott insulators.
Besides, the high pressure phase of FeO may have an important
implication in the earth science because FeO is believed to be one of the
major constituents of earth's deep mantle.

Under normal pressure and room temperature, FeO and MnO both take
the rock-salt (B1) structure with rhombohedral distortion along the $<111>$ direction.
This distorted B1 structure is called rB1 hereafter.
It was observed that FeO
undergoes a phase transition from the rB1 structure to a high pressure phase at
pressures above 70GPa~\cite{FeO1,FeO2}, whose crystal structure was
recently assigned to be the B8 by the analysis of
x-ray diffraction peak positions~\cite{Mao,Fei}.
On the analogy of almost all of the compounds with the B8 structure, a natural
idea for the B8 (NiAs) FeO may be such that Fe occupies the Ni site and
O the As site.  This structure is named nB8.  However, another
structure, which is named iB8, is possible by exchanging
the Fe and O positions~\cite{Cohen1}.
More recently, not only shock compression~\cite{Noguchi}
but also static compression~\cite{Kondo} experiments on MnO also showed
the existence of a possible high-pressure phase above 90GPa.
However, the crystal structure has not been determined yet.

In the present work, we have performed the first-principles density functional calculations
to give valuable hints to such problems as those mentioned above.
The electron-electron interaction is treated by the generalized gradient
approximation (GGA)~\cite{GGA} like in other similar
calculations~\cite{Cohen1,Sherman}.
However, the GGA is not powerful enough to treat the strong correlation
system such as Mott insulators.   Therefore, the GGA calculation
is supplemented by the LDA+U method~\cite{ldau} particularly in the low pressure 
regime where MnO and FeO are regarded as Mott insulators.

Highly converged total energy calculations are performed for different crystal
structures (B1, B2 (CsCl type), nB8, iB8) and spin structures (ferromagnetic(FM), AF)
for different volume (shown in Figure~1).
Surprisingly, the iB8 structure
with the AF ordering is the most stable for FeO among several
structures (including rB1) in the whole volume range.
First we pay attention to
the relative stability between the iB8 and nB8 structures in the
compressed volume range, where the GGA calculation will be reliable because
of the reduced $U/W$.  Clearly for FeO, the iB8 structure is more stable than
that of nB8.
At this stage, two fundamental questions are to be answered: 1) why is
the nB8 structure is realized rather than iB8 for most of
the transition-metal compounds with the B8 structure
(including the high-pressure phase of MnO, see Fig.~1(b))?;
2) what is special for FeO for the strong stability of iB8?
Anions around a transition metal ion form an octahedron in nB8
and a trigonal prism in iB8.
In the latter case, absence of the local inversion
symmetry about the transition metal ion reduces the strength of the
hybridization of the $3d$ orbitals with the oxygen $2p$ orbitals.
As the $p$-$d$ hybridization contributes to the stability of the
structure, this aspect favors the nB8 structure and will explain
the general feature that the nB8 structure is actually realized in most cases. 
However, the calculated energy level diagram 
shows existence of a well defined band gap between the conduction band and the valence band 
for the AF-iB8 structure of FeO 
in the whole volume range shown in Fig.~1(a), 
contributing to special stability of this structure.

A puzzling feature of Fig.~1(a) is that the iB8 structure
is significantly more stable than the rB1
structure even at equillibrium volume
at zero pressure.  This reflects the fact that the GGA cannot describe
the electronic structure of
Mott insulators properly: the GGA incorrectly makes FeO at normal pressure metallic.
In order to reproduce the correct ground state of FeO, we have to take account of
the local electron correlation and the spin-orbit interaction as well.
We adopt the LDA+U method with the LMTO (Linear-Muffin-Tin Orbital) basis~\cite{lmto}
to perform calculations with these ingredients.
A reasonable value of 4 eV as the effective Coulomb
interaction parameter $U_{\rm eff}$
makes the rB1 structure insulating and more stable than the iB8 structure
at normal pressure.  
On the other hand, for a compressed volume with experimentally determined 
lattice parameters ~\cite{Fei},
the same calculation still makes the iB8 structure most stable.
Therefore,
the puzzling feature of Fig.~1(a) is removed by the electron correlation effect.

According to the present calculation,
the high-pressure phase of FeO with the iB8 structure should be
insulating in contrast to the experimental observation of
metallic behavior~\cite{metal1,metal2}. There are two possible origins in this
disagreement.   Our calculation assumes the stoichiometric FeO, while
real samples contain about 5 \% Fe deficiency.  As the iB8 FeO is a
band insulator rather than a Mott insulator, itinerant carriers will be
doped by Fe deficiency.  Another possibility is related to the
magnetic disorder in the temperature range of the observed metallic behavior
($\sim 1000$ K).
According to the present calculation, even if the crystal structure is the iB8 type, 
the FM order makes the system
metallic, suggesting that the metallic state may be caused by the
loss of the AF order at such a high temperature range.

There is another strong evidence for the iB8 structure
as the high-pressure phase of FeO.  We have found that the intensity profile
of the observed x-ray diffraction pattern~\cite{Fei} can be reproduced only by
the iB8 type but not by the nB8 type.  To be more concrete,
we consider the relative intensity between (100)
and (101) peaks as an example.  Experimentally,
the latter is stronger than the former.
This feature is correctly reproduced only by the iB8 structure.

As for MnO, the present calculation shown in Fig.~1(b) suggests that
the most stable high-pressure phase will be the nB8 structure rather than
the B2~\cite{Noguchi} and iB8 structures.  Detailed comparison of the total
energies predicts that 
the FM-nB8 structure has the lowest energy
rather than the AF-nB8 structure,  
though the energy difference is rather marginal.

The present theoretical prediction can explain the recent
experiment~\cite{Kondo} on MnO in a high pressure
range ($> \, 120$ GPa) very well. The consistency of the assignment
of the experimental x-ray diffraction peaks by the nB8 structure
is demonstrated in Table~1.   Almost an exact fit of peak position 
and good agreement of the intensity profile can be
obtained except one peak for $d_{\rm exp}=1.844 {\rm \AA} \,$ which may
originate from some intermediate phase.   The intensity of this peak
is actually reduced after annealing.
The volume (=7.9 cm$^3$/mole) and the $c/a (=2.08)$ estimated by fitting the peak
positions are in good agreement with the present calculated results.

\begin{table}
\caption{Observed [8] and fitted x-ray diffraction patter of MnO at 137GPa after laser annealing.}
\begin{center}
\begin{tabular}{c|c|cccc|c}
\hline \hline
$d_{\rm exp}$$^{\dag}$ &$I_{\rm exp}$$^{\ddag}$ &$d_{\rm fit}$$^{\dag}$ (nB8) &h &k &l &$d_{\rm exp}$-$d_{\rm fit}$  \\  \hline  
2.534    &m        &2.538        &0 &0 &2   &-0.004       \\
2.110    &s        &2.114        &1 &0 &0   &-0.004       \\
1.955    &w        &1.952        &1 &0 &1   &0.003      \\
1.844    &s        &\multicolumn{4}{c|}{not coming from nB8} & \\
1.628    &s        &1.624        &1 &0 &2   &0.004       \\
1.218    &m        &1.220        &1 &1 &0   &-0.002        \\
1.099    &m        &1.100        &1 &1 &2   &-0.001        \\ \hline \hline
\end{tabular}
\end{center}
\dag\ $d_{\rm exp}$ and $d_{\rm fit}$ are experimental and fitted d-spacings respectively in unit of \AA. The fitted hexagonal unit cell has $a=2.441{\rm {\AA}}$ and $c=5.076{\rm {\AA}}$.

\ddag\ The relative intensities of the peaks are described as strong(s), medium(m) and weak(w).

\end{table}

Figure~2 shows the P-V relation for MnO obtained by the present
calculation and also by experiments.
Clearly, in the low pressure range, the sample is in the AF-rB1 phase.
The experimental data point around 120 GPa by shock compression
is located just in between the curves corresponding to the rB1 and nB8
structures, suggesting that the sample is in a mixed phase.  On the
other hand, the highest pressure data point obtained by the static
compression followed by laser annealing is just on the line
of the P-V curve for the nB8 structure.

We would like to make a brief comment on the $c/a$ value of the
nB8 of MnO and iB8 of FeO.  In the high-pressure phases, $c/a$ of
both systems exceeds 2.0, being unusually large compared with
the values for most of
other related systems.
However, we have found that $c/a$  is an increasing function of
$r_{\rm c}/r_{\rm a}$ with $r_{\rm a}\,\,(r_{\rm c})$
denoting the anion (cation) ionic radius and that the $c/a$ values
of the MnO and FeO are on the extrapolated line of this general trend.

Finally, as one of the important consequences of the
present analysis, strong stability of the iB8 phase of FeO
may imply strong tendency of the
incorporation of oxygen in Earth's core.


{\bf Methods}

{\bf Pseudopotential calculation.} The plane-wave basis pseudopotential method is used to perform the structural optimization efficiently.  The $2p$ states of oxygen and $3d$ states of Mn and Fe are treated by the Vanderbilt ultrasoft pseudopotential~\cite{PP}. The cutoff energy for the plane-wave expansion is $4.9 \times 10^2$ eV, and sampling of k-points gives absolute total energy convergence to better than $10^{-2}$ eV/fomular unit ($\sim 1.0$ KJ/mole). Moreover, note that the convergence in the relative energy between different structures is generally order of magnitude much better than the convergence in the absolute total energy. The equation-of-state parameters obtained by the present calculations agree well with those obtained by other similar calculations~\cite{Sherman,Cohen2}. High credibility of the present approach has been well confirmed.

{\bf LDA+U method.} The interaction between localized $d$ electrons is explicitly taken into account through the screened effective Coulomb interaction parameter $U_{\rm eff}$.

{\bf Estimation of the $p$-$d$ hybridization.} The reduction in the $p$-$d$ hybridization in the iB8 structure compared with that in the nB8 structure was confirmed in the present calculation by estimating the energy separation between the centers of the $p$ and $d$ bands.



{\bf Acknowledgements} The authors thank Professor Y. Syono, 
Professor T. Yagi and Dr. T. Kondo
for many valuable comments and for providing us with their
experimental data before publication.  The present work
is partly supported by NEDO.

\vspace{2cm}

Correspondence and requests for materials should be addressed to Z. Fang 

(email: zfang@jrcat.or.jp)

\clearpage
\onecolumn
\begin{figure}
\centering
\epsfile{file=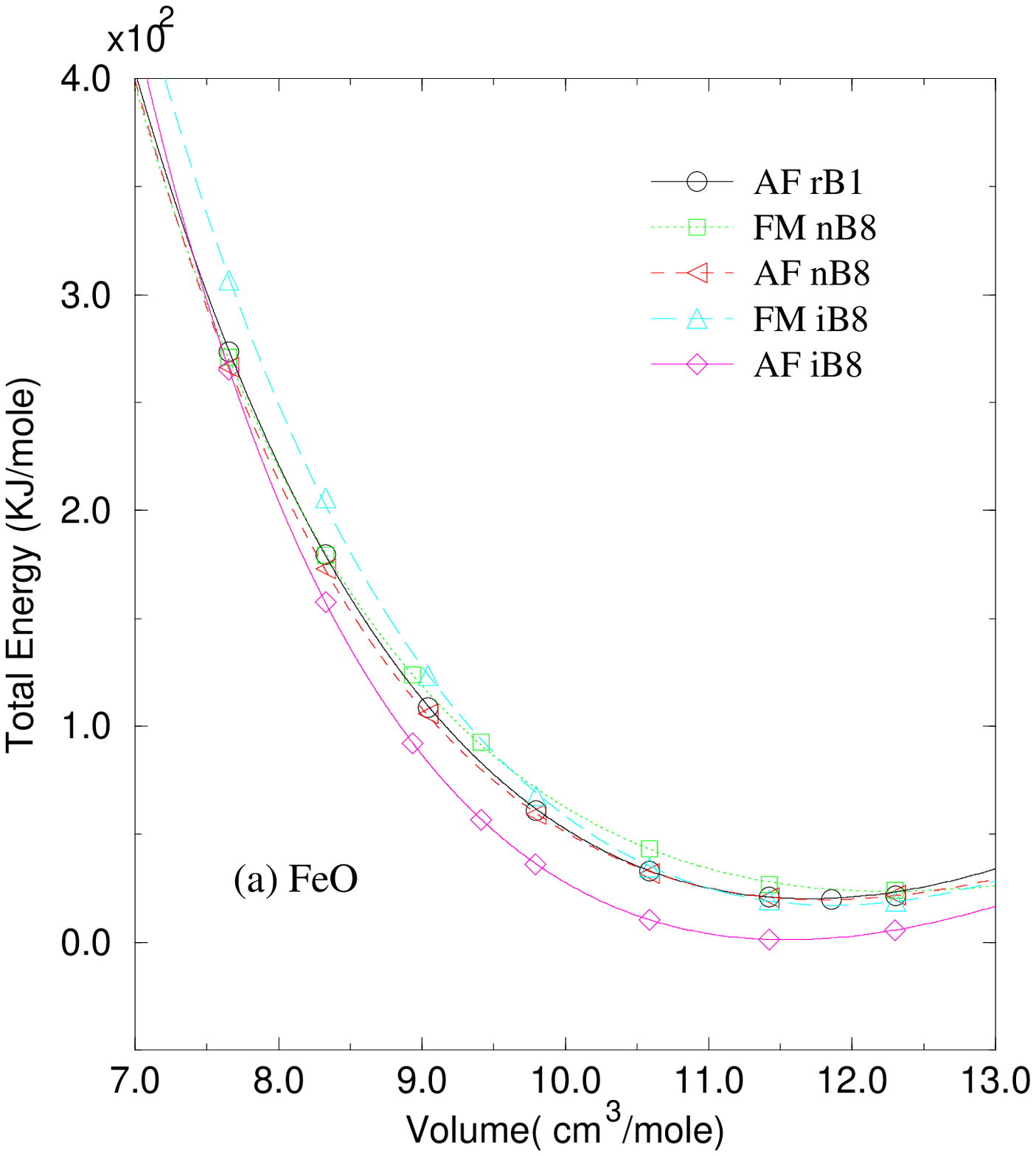,vscale=0.6,hscale=0.6}
\end{figure}

\begin{figure}
\centering
\epsfile{file=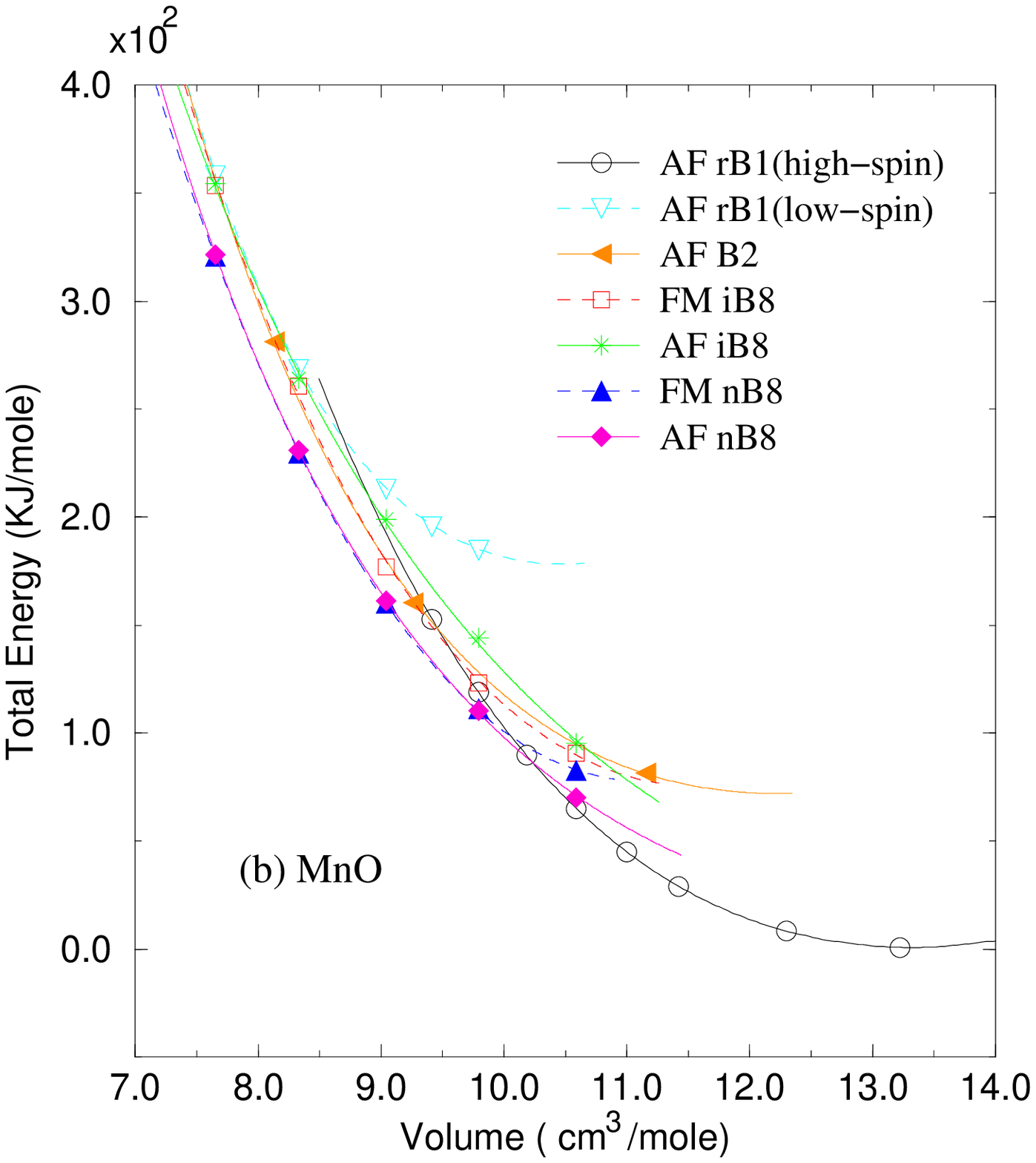,vscale=0.6,hscale=0.6}
\caption{The total energies of FeO (a) and MnO (b) as a function of the volume. The least-squares-fitted curves to the Murnaghan's equation of state [19] are shown.}
\end{figure}

\clearpage
\begin{figure}
\centering
\epsfile{file=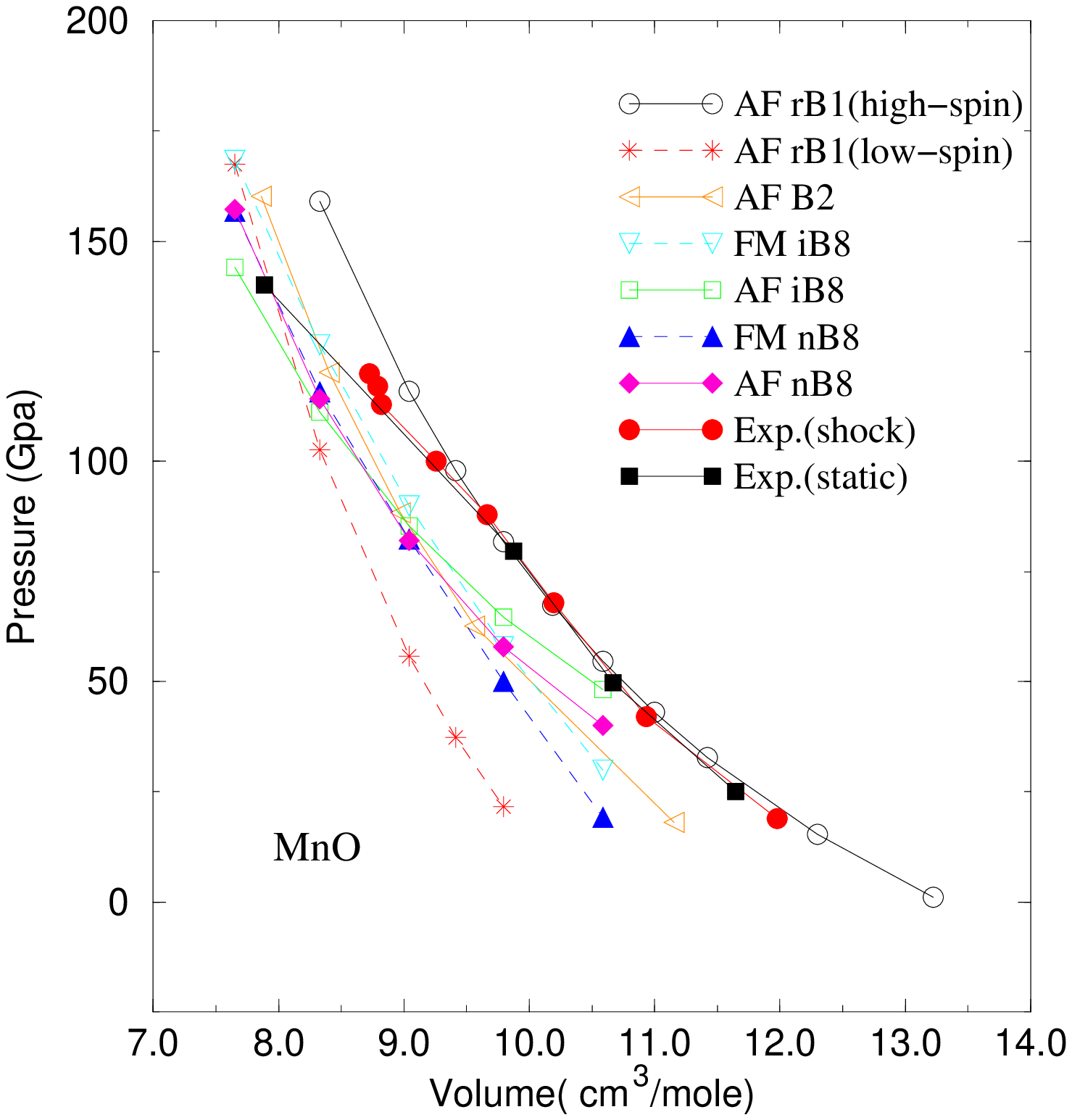,vscale=0.6,hscale=0.6}
\caption{The Pressure-Volume curves for different phases of MnO.}
\end{figure}

\end{document}